\documentclass[10pt, reprint, aps, amsmath,amssymb,twocolumn]{revtex4-1}

\usepackage{epigraph}
\usepackage{graphicx}
\usepackage{dcolumn}
\usepackage{bm}
\usepackage[font=scriptsize]{caption}
\usepackage{booktabs}
\usepackage{subcaption,graphicx}

\begin{document}

\title{Characterizing Germanium Junction Transistors}

\author{Luciano da F. Costa}
\email{ldfcosta@gmail.com, luciano@if.sc.usp.br}
\affiliation{S\~ao Carlos Institute of Physics, IFSC-USP,  S\~ao~Carlos, SP,~Brazil}

\date{\today}

\begin{abstract}
Transistors have provided the basis of modern electronics.  Being relatively intricate devices,
and often exhibiting intense parameter variation, this type of electronic devices
has motivated much research interest especially regarding their characterization and modeling.
In this work, we apply a recently reported modeling methodology, based on the Early effect, for
characterizing new old stock NPN and PNP small signal germanium junction transistors and 
comparing them to more modern silicon bipolar junction devices. The Early approach is
special in the sense that its two parameters, namely the Early voltage $V_a$ and a proportionality
parameter $s$, are fixed and independent of transistor operation.   Remarkable results are
obtained, including the fact that the four considered groups, namely NPN and PNP germanium
and silicon devices, occupy mostly non-overlapping regions in the Early parameter space,
with PNP devices presenting larger parameter variability in both cases.  Surprisingly, the considered
germanium devices exhibited smaller parameter variability than observed for the silicon counterparts.  
When mapped into the more traditional space defined by the current gain and output resistance parameters,
the four transistor groups exhibited much larger overlaps and yielded a clustering structure much
less organized than allowed by the Early mapping. This result suggest that the Early representation
of transistors is more compatible and inherently related to the structure of amplifying devices such
as those considered in this work.   In addition, it was verified that the center of mass of each of the
NPN-PNP pairs of germanium and silicon devices are crossed by respective $\beta$-isolines for
gains of 130 and 250, respectively.  Germanium devices were also characterized as having smaller
output resistance and smaller magnitudes of Early voltage.  
\end{abstract}

\maketitle

\setlength{\epigraphwidth}{.49\textwidth}
\epigraph{\emph{``The only thing we know about the future is that it will be different.''}}{Peter F. Drucker}

\section{Introduction}

It is not so often realized that most of modernity has been supported, and to some extent even defined,
by \emph{electronics}.   Though the analogue is frequently said about \emph{computing}, this important area 
itself relies strongly on electronics.  As the development of modern electronics started with the transistor invention in 
1947~\cite{riordan:1997}, the above sentence could be rephrased as \emph{modernity, ultimately, being induced 
by the transistor}.  As a matter of fact,  any cell phone or personal computer, not to say the myriad of other electronic 
devices and systems that permeate our lives, rely critically on millions of transistors -- it remains an interesting question 
to estimate the number of transistors involved in the Internet.  Though most of these devices employ silicon as  basic 
semiconductor  material, the first transistors employed germanium, which was dominant from 1947 to the mid 50's~\cite{riordan:1997} and
used until the 60's and even 70's.  While MOSFET technology mostly predominates in modern analog electronics,
point contact, grown junction, alloy junction and surface barrier were employed from 1947 to mid 50's.  

Both germanium (Ge) and silicon (Si) are semiconductor materials, characterized by increase of conductivity with temperature
as free carriers are transferred from the valence to the conduction bands (e.g.~\cite{parker:2004,sze:1969,jones:1995}).
However, as summarized in Table~\ref{tab:feats}, these two semiconductor materials have quite
distinct physical and electronic properties.  In particular, the substantially larger mobility of germanium has great influence on the 
electronic properties of respectively derived transistors.  Though this would represent, in principle, an advantage for germanium 
devices, there are some shortcomings potentially constraining the electronic applications of these devices relatively to silicon 
counterparts.  These include the fact that more free electrons are available at room temperature in germanium than in silicon, 
implying considerably smaller current cut-off  ($I_{CBO}$) for silicon devices.   Germanium-based transistors are also characterized by
smaller peak inverse voltage ($PIV$) than silicon devices, averaging $350V$ and $1000V$, respectively.  In addition, the typical 
working temperature for germanium is only about $70^o$ for germanium, being much larger (about $160^o$) for silicon.  Such issues 
contributed to the progressive shift from germanium to silicon as the basic semiconductor material adopted for industrialized transistors.  

\begin{table}
  \centering
     \begin{tabular}{| l || c | c | c | c | c | c | c | c | c | c}
        \hline
        Property & Silicon & Germanium    \\  \hline \hline
        Electron mobility $(cm^2/(Vs))$ & 1500 &  3900 \\ 
        Hole mobility $(cm^2/(Vs))$ & 470 & 1900  \\ 
        Electron effective mass & 300 & 500  \\ 
        Band gap $(eV)$ & 1.1 & 0.67   \\ 
        Dielectric constant & 11.9  & 16.0 \\
        Melting point $(^oC)$ & 1415 & 937   \\  \hline
        $I_{CBO}$  $(\mu A)$ & $\approx 0.05$ & 2   \\
        $PIV$ $(V)$ & 1000 & 350   \\
        Typical working temperature $(^oC)$ & 160 & 70   \\       \hline
      \end{tabular}
    \caption{Some of the main contrasting physical and electronic properties of germanium and silicon.}
    \label{tab:feats}
\end{table}

Except for $I_{CBO}$, the germanium relative limitations are more critical only for given applications and circumstances,
remaining a viable choice for many important usages, such as in many small signal and low power linear circuits, especially
when special attention is given to temperature management and circuit design.  So, germanium remains a potentially 
interesting choice in transistor electronics.  Some important issues that could impact on the eventual use of germanium devices
in linear electronics concern their electronic parameters (e.g.~\cite{stewart:1956, jaeger:1997,sedra:1998,boylestad:2008,jaeger:1997}), 
such as  the current gain $\beta$ and output resistance $R_o$, as well as
frequency features such as $f_T$.   Relatively little can be found in the literature regarding the more systematic characterization of
real-world germanium devices.  Actually, even the available data 
is often of relatively limited assistance because both $\beta$ and $R_o$ tend to vary largely with $I_C$ and $V_C$ during normal 
circuit operation, while the given parameter values are typically limited to averages or specific to given operation points.  
A substantially more comprehensive characterization of the electronic properties of a transistor can be achieved
by considering the whole region of operation in the  $V_C \times I_C$ characteristic space, where 
$V_C$ and $I_C$ stand for the collector voltage and current, respectively.
In addition, even when taken from the very same lot, transistors tend to present largely varying parameters, 
implying several devices to be considered so as to achieve statistical significance.

The fact that germanium transistors have some potentially interesting physical features, especially higher mobilities, while
their other limitations do not completely undermine their applications in many practical circuits and circumstances, 
entails the interesting prospect of verifying, in a more systematic and comprehensive way, how germanium transistors 
behave electronically.  In particular,  it would be interesting to compare small signal germanium and 
silicon devices, so as to try to identify their respective relative main pros and cons.  These issues motivate the present work.  
This is a timely endeavor because, with the with the advances in instrumentation and 
information technology, more systematic, stable and comprehensive approaches to transistor characterization have 
become available.  In particular,  we resource to the Early effect-related method reported 
in~\cite{costaearly:2017,costaearly:2018}, as well as the systematic approach to characterization of transistors
described in~\cite{costafeed:2017}.

The application of these methodologies, while considering new old stock germanium junction transistors, led to several
remarkable results, which are presented and discussed in this work.   Of particular relevance is the fact that the
distribution of the devices resulted much more definite and organized than that obtained in the more traditional and
electronically intuitive space defined by the current gain $\beta$ and the collector, or output resistance $R_o$.  This 
important result paves the way to more systematic characterization studies aimed at mapping several types of amplifying
devices in the Early space, so as to assist the analysis and design of electronic devices and circuits.

This article starts by revising the theoretic-experimental-numeric Early characterization 
approach~\cite{costaearly:2017,costaearly:2018} and proceeds
by presenting and discussing the results obtained with respect to three types of germanium junction transistors as well as
a comparison with results previously~\cite{costaearly:2018} obtained for 7 complementary types of silicon transistors.

\section{Early Modeling and Characterization}

\begin{figure}[h!]
\centering{
\includegraphics[width=8cm]{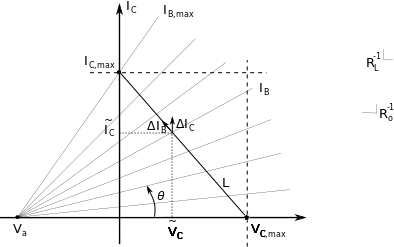}
\caption{The geometrical set-up of Early modeling, including the chosen region of operation (defined by $0 \leq I_C \leq _{C,max}$ 
and $0 \leq V_C \leq V_{C,max}$), as well as the load line defined by the load resistance $_L$ and $V_{C,max}$.  The
lines radiating from $(V_a$,0) are henceforth called \emph{isolines} as they are defined and respectively indexed by constant values of  
$I_B$. }
\label{fig:earlyg}}
\end{figure}

In 1952, an interesting electronic effect taking place in junction semiconductors was reported by J. M. Early, in which the
width of the charge carrier portion of the base varies with the base-collector voltage.  In typical transistor configurations,
this implies that, for a fixed current base $I_B$, the collector voltage $V_C$ will not be zero for $I_C=0$, but rather 
would take a negative value $V_a$ (the Early voltage), were it possible to cross over to negative values of $V_C$ .
This is illustrated in Figure~\ref{fig:earlyg}, which also shows the chosen region of operation (defined by $0 \leq I_C \leq _{C,max}$ 
and $0 \leq V_C \leq V_{C,max}$) and the load line defined by a chosen load resistance $_L$ and $V_{C,max}$
considering a simplified common emitter configuration in which the load is attached between the voltage supply $V_{CC}$ and
the collector.  \emph{It should be taken into account that, for simplicity's sake, all voltages and currents for the PNP devices
are taken in absolute values in this work.}

The angle $\beta$ has been experimentally found~\cite{costaearly:2017,costaearly:2018}, at least for the devices considered in that
work, to be directly proportional to the base current,  i.e. $\theta = s I_B$.  The Early voltage $V_a$ and the proportionality 
parameter $s$ constitute the two parameters of the Early modeling.  This linear relationship turns out to be of great importance
for deriving and applying the Early model, as it considerably simplifies the respective calculations and contributes to the simplicity of the 
approach and for the derivation of several analytical relationships.  Observe that the Early model is also special in the sense that its two just
mentioned parameters do not vary with either $I_C$ or $V_C$, as it happens with the large majority of the more traditionally
adopted  junction transistors models. The more commonly adopted parameters known as  current gain ($\beta$) and output 
resistance ($R_o$), also illustrated in Figure~\ref{fig:earlyg}, are defined  (e.g.~\cite{stewart:1956} respectively as:

\begin{eqnarray}
   \beta = \frac{ \partial{I_C}}{\partial{I_B} }\bigg|_{V_C}   \\
   R_o =  \frac{\partial{V_C}}{\partial{I_C}} \bigg|_{I_B}
\end{eqnarray}

In this more traditional parametrization, we have that both $\beta$ and $R_o$ are functions of
the two collector variables, i.e. $\beta = \beta(V_C,I_C)$ and  $R_o = R_o(\tilde{V_C},\tilde{I_C)}$.  So, at a given point
$(\tilde{V_C},\tilde{I_C})$, $\beta$ needs to be approximated as $\beta \approx \Delta I_C(\tilde{V_C},\tilde{I_B}) / \Delta 
I_B(\tilde{V_C},\tilde{I_B})$, while $V_a$ (and $s$)  remains constant along the whole $(V_C,I_C)$ space.  Still, $\beta$ and $R_o$
are also required, because they correspond to more intuitive electronic properties of the transistors.  Fortunately, it is possible to
derive relationships~\cite{costaearly:2018} between the Early parameters and the values of $\beta$ and $R_o$, both averaged along the
region of operation  $0 \leq I_C \leq _{C,max}$ 
and $0 \leq V_C \leq V_{C,max}$, which are given~\cite{costaearly:2018} as 

\begin{table*}
\begin{eqnarray} \label{eq:rels}
	\langle \beta \rangle =   \frac{s}{6 V_{C,max}} \left[   2 I_{C,max}^2 ln \left( \frac{V_a-V_{C,max}}{V_a} \right) + 3 V_{C,max}^2 - 6 V_a V_{C,max}  \right]  \label{eq:betar} \\
        \langle R_o \rangle = \frac{\left( V_{C,max}^2 - V_{C,max} V_a \right) }{V_{C,max}I_{C,max}} ln \left( \frac{I_{C,max}} {I_{C,min}} \right)  
                  \label{eq:Ror}
\end{eqnarray}
\end{table*}  

where $V_{C,max}$ is the maximum collector voltage considered (in this work, $V_{C,max}=8V$), $I_{C,min}$ is the minimum
collector current needed to be taken into account so as to avoid divergence in Equation~\ref{eq:Ror} (here, $I_{C,min} = 1mA$),
and $I_{C,max}$ is the maximum considered collector voltage (henceforth taken as $I_{C,min} = 15mA$).  In this
way, the relative advantages of the two parametrizations can be combined.

The Early parameters $V_a$ and $s$ of a real-world transistor can be numerically estimated by applying the following three 
steps~\cite{costaearly:2017,costaearly:2018}: (i) the values of $V_C$, $I_C$, $V_B$ and $I_B$ are obtained experimentally by scanning
the device with a succession of $V_{CC}$ values (note that other schemes can be used); (ii) a Hough transform accumulation (or voting) 
scheme is employed to identify the point in the $(V_C,I_C)$ space where the isolines (obtained
by linear regression) intercept, leading to the estimation of $V_a$; and (iii) linear regression is applied in order to obtain $s$
from the parameters $I_B$ and $\theta$, the latter corresponding to the tangent of the angular coefficient of the respective
isolines.  

The above outlined methodology has been applied with encouraging success for the experimental characterization of BJT NPN-PNP 
complementary pairs~\cite{costaearly:2018}, yielding several interesting results including the identification of two almost non-overlapping 
groups defined respectively by these two types of devices, each with well-defined specific electronic properties.  In addition, all 
the considered devices were found to populate a relatively 
narrow curved band in the Early space, with $\langle \beta \rangle$ varying from 100 to 400.  The main distinction between the PNP 
and NPN groups was that the former presents intrinsic larger parameter variation, as well as smaller magnitudes of $V_a$ and 
larger values of $\langle s \rangle$.  A prototypical Early space was outlined in ~\cite{costaearly:2018}, which provides a
reference for comparison with other families or types of transistors and other amplifying devices.   In the current work,
the group of Germanium semiconductors is included into that Early prototype space.

\subsection{Results for Germanium Devices}

The experimental data considered in the present work derives from three types of small signal germanium junction transistors,
2 NPN and 1 PNP.   All used germanium transistors are of alloy junction type. Ten samples of each were selected from homogeneous 
respective new old stocks.  The acquisition,
pre-processing and analysis of the respectively derived signals followed the methodology described 
in~\cite{costaearly:2017,costaearly:2018}.

\begin{figure}[h!]
\centering{
\includegraphics[width=9cm]{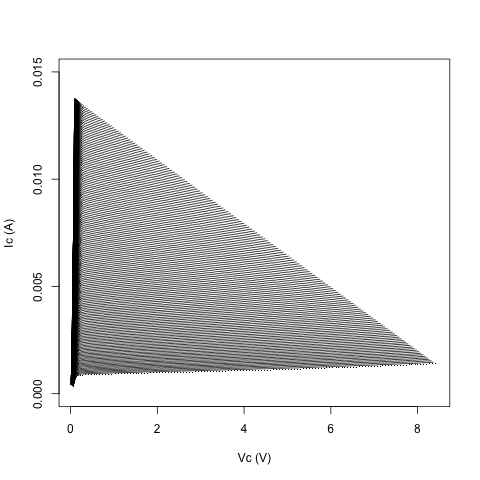}
\caption{Example of isolines obtained for one of the NPN germanium transistors.  These isolines are shown as 
acquired, without smoothing or resampling.  The noise has been verified to be in the order, or smaller, than the
resolution of the 12-bit ADCs, as a consequence of special care taken in the development of the customized
acquisition system.}
\label{fig:raw}}
\end{figure}

Figure~\ref{fig:raw} illustrates one of the obtained sets of $I_B-$indexed isolines as acquired, without any
smoothing or pre-processing, which indicates a good signal to noise level.  Observe the increasing slope
of the isolines as $I_B$ goes from $0$ (bottom) to $I_{B,max}$ (top), which is an intrinsic characteristic of the
Early effect.  Next, each of these sets of isolines is 
slightly smoothed through an average filter (11 points wide) and resampled so as to have the same number of
isolines in all cases~\cite{costaearly:2017,costaearly:2018}.

\begin{figure}[h!]
\centering{
\includegraphics[width=8cm]{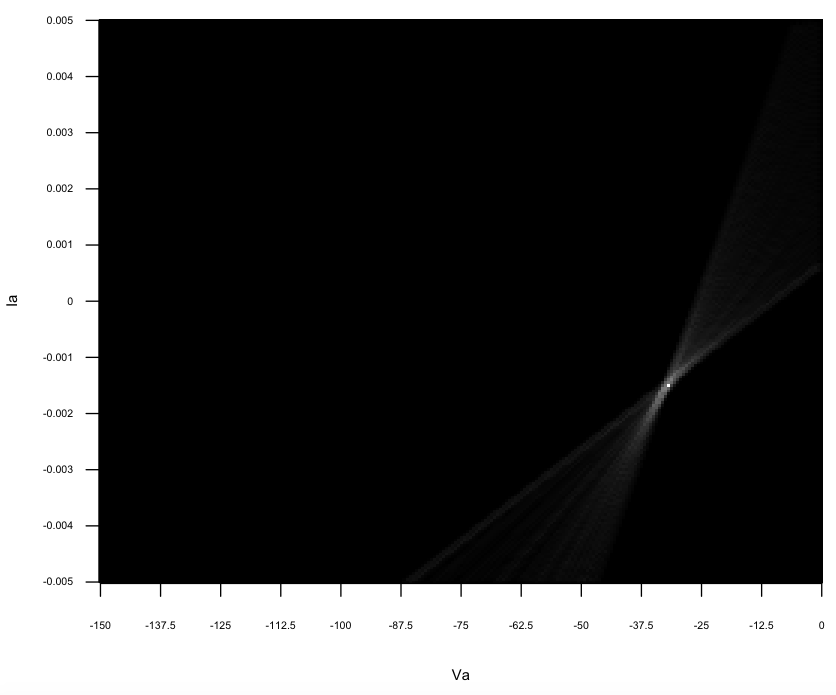}
\caption{The accumulator array obtained from equispaced version of the isolines for the transistor
in Figure~\ref{fig:raw}.  A well-defined peak is defined by the intersections of those isolines in the
accumulator array.  The $V_a$ coordinate of the intersection point is taken as the most
likely estimation of the Early voltage for that transistor.}
\label{fig:acc}}
\end{figure}

The Hough transform-inspired voting (or accumulating) scheme used for numeric estimation of $V_a$
typically resulted in sharp, well-defined peaks such as that illustrated in Figure~\ref{fig:acc}.  Observe
that only a region of the $(V_a,s)$ space, corresponding to where the $I_B$ isolines tend to intersect,
needs to be considered in the Hough mapping and voting. The $V_a$
coordinate ($x$-axis) of the peak is taken as the most likely value of $V_a$ for each transistor. Observe
that the peak of isolines intersections in the $(V_a,s)$ space does not occur exactly on the $V_a$ axis,
with a small offset in the order of one $mV$ being obtained in the $I_C$ peak coordinates.  The use of the
Hough transform-inspired voting allows eventual isolines diverging too much from the others not to
be considered.

\begin{figure}[h!]
\centering{
\includegraphics[width=8cm]{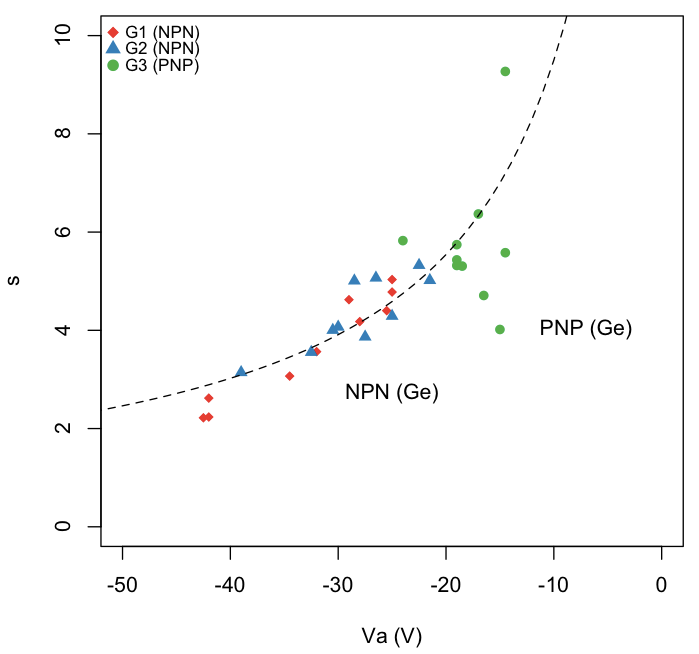}
\caption{Mapping of the three considered groups of germanium junction transistors in the Early space.
The two NPN groups, identified as $G1$ ad $G2$, present great overlap, but differ substantially
from the PNP group $G3$. The isoline for $\langle \beta \rangle = 130$, also represented by the
dashed curve in this figure, passes very near the center of mass (average) of the NPN and PNP groups.}
\label{fig:Earlysp}}
\end{figure}

Figure~\ref{fig:Earlysp} presents the obtained distributions of germanium NPN and PNP junction transistors
in the Early space $(V_a,s)$.  The two NPN types, namely $G1$ and $G2$ resulted with very similar
Early parameters, therefore exhibiting great overlap in the Early space in Figure~\ref{fig:Earlysp}.  This
means that these two types of NPN germanium transistors have very similar electronic properties, at least
considering the adopted samples/configurations and non-reactive loads.  The PNP group, identified as $G3$ 
in this figure, resulted with distinctively higher $V_a$ values (smaller $V_a$ magnitudes), and also larger 
values of the proportionality parameter $s$ than the devices in the two NPN groups.  The two groups have 
their centers of mass (averages) very near to the isoline defined for $\langle \beta \rangle = 130$, which is shown by 
the dashed curve in this figure.  

\begin{figure}[h!]
\centering{
\includegraphics[width=8cm]{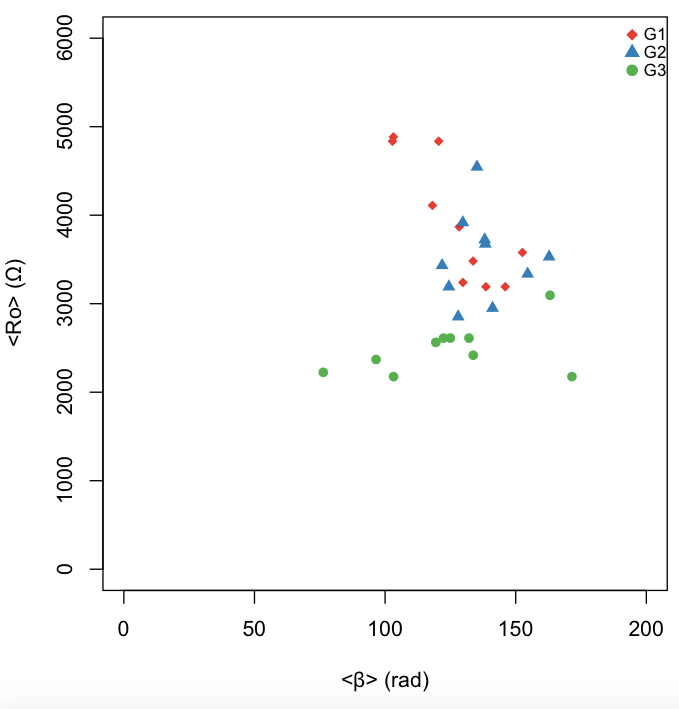}
\caption{The mapping of the three group of germanium junction transistors in the more traditional space
induced by parameters $\langle \beta \rangle$ and $\langle R_o \rangle$. The NPN and PNP groups result
nearer one another in this space.}
\label{fig:betaRo}}
\end{figure}

The scatterplot derived by the adopted Early methodology for the more traditional and electronically intuitive
parameters $\langle \beta \rangle$ and $\langle R_o \rangle$ is depicted in Figure~\ref{fig:betaRo}.  As in the
Early mapping, the two NPN groups largely overlap one another, and are also less separated from the PNP than in
that other space. Similar levels of parameter variations are observed in this space for any of the three types of
germanium transistors.  In addition, the PNP devices yielded smaller $R_o$ values.

\subsection{Comparison with Silicon BJTs}

\begin{figure*}[h!]
\centering{
\includegraphics[width=18cm]{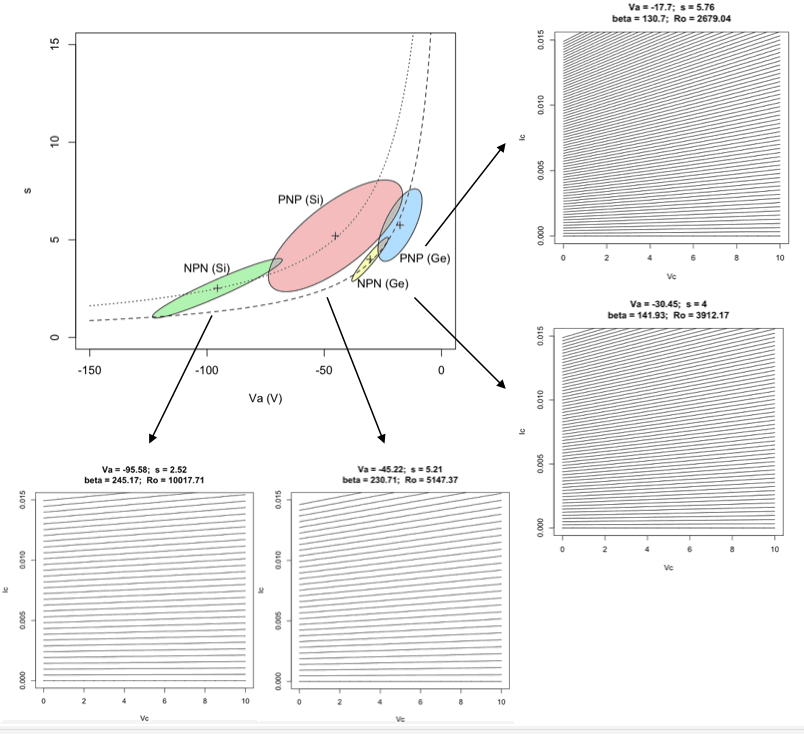}
\caption{The considered four prototypical groups of devices --- NPN (Ge), PNP (Ge), NPN (Si) and PNP (Si)  --- represented 
in the Early space. 
The respective dispersions are reflected in the ellipses, while the means are represented as ``$+$''.  The isolines corresponding to 
$\langle \beta \rangle = 250$ and $\langle \beta \rangle = 130$ are also given (dotted and dashed lines, respectively).  The prototypical
$(V_C,I_C)$ characteristic surfaces for each group are illustrated by respective separated graphs. The overlapping regions are
identified in gray.}
\label{fig:prots}}
\end{figure*}

It would be particularly interesting if the electronic characteristics obtained for germanium junction transistors could be compared 
with those of more modern silicon bipolar junction transistors -- BJTs.  Fortunately, this can be done by using recently 
obtained results~\cite{costaearly:2018} regarding the latter
type of transistors.  A total of 14 types of small signal BJTs (7 NPN and 7 PNP), each represented by 10 samples, were characterized
in that work by using the same Early approach adopted here.  Figure~\ref{fig:prots} presents the mean and dispersion (Mahalanobis ellipses~\cite{johnson:2002}) obtained for the four groups --- NPN (Ge), PNP (Ge), NPN (Si) and PNP (Si) -- in the $(V_a,s)$
Early space.  Each ellipse was obtained from the respective covariance matrices by considering unit Mahalanobis 
distance~\cite{johnson:2002}, specifying the same number of samples to comprised, in the average, inside each respective ellipse.
The means of each group are marked by the ``$+$'' sign.  The isolines corresponding to $\langle \beta \rangle = 250$ and
$\langle \beta \rangle = 130$ are also represented by the dotted and dashed lines, respectively.  The prototypical characteristic
surfaces respectively induced in the $(V_C,I_C)$ transistor operation space are also shown in the separated plots, providing
a direct graphic representation of the typically expected electronic properties for each of the considered groups.

Several remarkable results can be inferred from Figure~\ref{fig:prots}.  First, we have that the NPN-PNP groups obtained
for germanium look like the silicon pairs scaled so as to have smaller $V_a$ magnitude and slightly higher values of $s$.     
Interestingly, the elongation of all groups tend to be mostly aligned with the tangent of the respective
$\beta$ isolines at their respective centers of mass.  More specifically, the germanium and silicon pairs have, in the average,
respectively $\langle \beta \rangle \approx 250$ and $\langle \beta \rangle \approx 130$.  Thus, the silicon devices tend to have, in the 
average, twice as much current gain and smaller $V_a$ magnitude values, implying larger $R_o$ magnitudes.  The average 
$R_o$ values obtained for the germanium NPN and PNP transistors were smaller than those obtained for the two silicon groups.  
PNP devices tended to present larger parameter variability
for both germanium and silicon, but germanium transistors were -- surprisingly -- characterized by smaller  absolute parameter 
variations than those observed for silicon counterparts.

\begin{figure}[h!]
\centering{
\includegraphics[width=8cm]{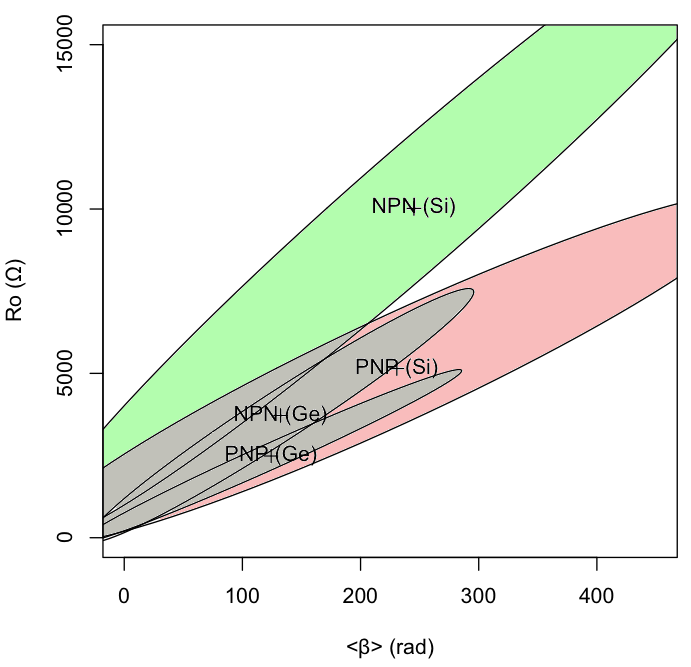}
\caption{The considered four prototypical groups of devices --- NPN (Ge), PNP (Ge), NPN (Si) and PNP (Si)  --- represented in 
terms of the more traditional and electronically intuitive parameters $\langle \beta \rangle$ and $\langle R_o \rangle$. 
The overlapping regions are identified in gray. Only two regions result without overlap, namely those corresponding to silicon
transistors but, even so, the region corresponding to their PNP group is severely constrained by overlaps.  It should
be observed that the orientation of the ellipses in this figure appear not to be visually aligned with the respective groups is a 
consequence of the fact that the $y-$axis values are much larger than those at the $x-$axis.}
\label{fig:prots_betaRo}}
\end{figure}

Figure~\ref{fig:prots_betaRo} depicts the four transistor groups in the more traditional parameter space defined by $\langle \beta \rangle$
and $\langle R_o \rangle$, also including the respective unit Mahalanobis ellipses.  Unfortunately, the relative properties of the considered
transistor groups can hardly be discerned in this space, as many ellipses overlap strongly.  In addition, the relative
features of the transistor groups, so evident in Figure~\ref{fig:prots}, can hardly be inferred from this parametric representation. 
This striking contrast of results can be understood as corroborating the efficacy and naturality of the Early space with respect to the more 
traditional parametrization induced by  $\beta$ and $R_o$.   The latter space, however, is underlain by more intuitive electronic
interpretation, especially when analyzing/designing circuits.  So, it may  be interesting to consider the twin representation of transistors 
in the Early as well as in more traditional parametric spaces, the former being more discriminative and inherently compatible
with the geometry of transistor operation and the latter being electronically more intuitive.

\section{Concluding Remarks}

The advances of electronics since the development of the transistor have been so
impressive and continuous that the critical importance that germanium devices
played from the 50's to the 60's tends to be overlooked.  The present
work reported a brief excursion into the origins of modern electronics.  More specifically,  
new old stock germanium junction transistors were characterized by a recently introduced
theoretic-experimental-numerical approach based on the Early effect, as well as on image 
processing/analysis and pattern recognition concepts and methods.  Two types of NPN 
and one type of PNP new old stock small signal germanium transistors were scanned
electronically and the obtained signals were used to derive the respective characterization
in the Early, as well as in the more traditional and electronically intuitive 
$(\langle \beta \rangle, \langle R_o \rangle)$ space. 

Remarkable results were obtained with respect to both the main reported investigations,
namely the characterization of germanium junction transistors and their comparison with
more modern silicon BJT devices.  Perhaps the principal contribution of this work is the
characterization of the considered four groups of devices --- NPN (Ge), PNP (Ge), NPN (Si) and PNP (Si) ---
in both the Early and the more traditional $(\beta,R_o)$ spaces.  This characterization was performed
in terms of the center of mass (average) and dispersion (Mahalanobis distance, derived from the respective
covariance matrices), as well as $\langle \beta \rangle$-indexed isolines.  A surprisingly well-defined
distribution of the transistor types resulted in the Early space, with the NPN devices exhibiting greater
parametric variation for both silicon and germanium.  Yet and remarkably, unlike it is sometimes believed,
the old germanium devices presented substantially smaller parameter variation than the considered modern
silicon devices.  The four groups of transistors also resulted remarkably separated, with little overlap between
them, indicating that,  at least for the adopted devices, these
groups have well-defined specific characteristics.   Germanium devices tended to have smaller $V_a$ magnitude
(hence smaller $R_o$), while yielding $s$ values comparable to those of silicon devices, and $\langle \beta \rangle$ 
was found to be about half of their silicon counterparts.   

At the same time, the mapping of the four groups of transistors in the more traditional and electronically intuitive
$(\langle \beta \rangle, \langle R_o \rangle)$ space yielded largely overlapping groups to a level that 
substantially undermines the identification of interesting relationships observed for the Early mapping.
It is believe that, because of their complementary features, these two spaces could be used jointly while
analyzing and designing devices and circuits.

It should be reminded that the reported results are specific to the devices, methods and configurations adopted
in this and previous works~\cite{costaearly:2018}.  In particular, a limited number of new old stock alloy junction transistors
were used, so that distinct results can be eventually obtained for other technologies (e.g.~grown junction) and even 
for other models of alloy junction devices.  Additional research is needed to complement the characterization 
under other circumstances and for additional device types.  In addition, it is important to bear in mind that the 
reported results were derived from new old stock
devices from the 50's and 60's, so that it becomes difficult to identify if the obtained characteristics 
derive from germanium material itself or from the fabrication methods and technology used in the 50's and 60's.  
It remains an interesting question to verify how modern junction germanium transistors would compare to the 
here obtained results for new old stock.   It would also be worth investigating further the influences of the 
features observed for the considered devices in typical circuit applications, including the determination of
gains, input and output resistances, distortion, and other properties.  Of particular interest would be to 
perform AC analysis of the devices and circuits, as the present work was limited to DC estimation of the
transistor features, as it is known that reactive components of transistors can strongly influence AC circuit
operation.    Yet, the features obtained for the considered new old stock germanium transistors are interesting themselves,
especially in the sense that they tend to populate areas of the Early space not well covered by silicon devices, therefore
providing valuable design alternatives regarding distinct $V_a$ and $s$ features.  As a matter of fact, it would be
of great potential value to have prototypes of additional devices (such as optocoupler, high frequency and high
power transistors, MOSFET, silicon-germanium heterojunction, Darlington, etc.) incorporated into the obtained Early 
space, so as to achieve a kind of 
general atlas that could be used to assist electronic analysis and design.   A possible reason why such 
maps have not become widely available is the fact that the parametric mapping of devices into more traditional spaces,
as hinted by the here considered and reported cases, tend to imply larger overlap between the groups.  The 
neat separation obtained in the Early space seems to substantiates this approach as providing a more natural and inherently 
effective means for mapping and studying junction transistors, as well as other similar devices.

\vspace{0.7cm}
\textbf{Acknowledgments.}

Luciano da F. Costa
thanks CNPq (grant no.~307333/2013-2) for sponsorship. This work has been supported also
by FAPESP grants 11/50761-2 and 2015/22308-2.
\vspace{1cm}

%\nocite{*}

\bibliography{mybib}
\bibliographystyle{plain}
\end{document}